\documentclass[12pt,a4paper]{article}
\usepackage{a4wide}
\usepackage{amsmath}
\usepackage{amssymb}
\usepackage{epsfig}
\usepackage{subfigure}
\usepackage{exscale}
\usepackage{float}
\usepackage{bbm}
\usepackage[numbers,sort&compress]{natbib}
\usepackage{pst-plot, pstricks,pst-math}
\usepackage{fancybox,amssymb,color}
\usepackage{graphicx}
\usepackage{pstricks, color, graphicx, epsfig, psfrag}
\usepackage{amsfonts,amsmath,amssymb,slashed}
\usepackage{dsfont}
\usepackage{bbm,bm}
\usepackage{fancyhdr, a4wide}
\usepackage[english]{babel}

\makeatletter
\@addtoreset{equation}{section}
\makeatother
\renewcommand{\theequation}{\thesection.\arabic{equation}}

\title{Cubic Ideal Ferromagnets at Low Temperature and Weak Magnetic Field}

\author{Christoph P.\ Hofmann \\ \\
\normalsize {Facultad de Ciencias, Universidad de Colima} \\
\vspace{0.3cm}
\normalsize {Bernal D\'iaz del Castillo 340, Colima C.P.\ 28045, Mexico} \\}

\begin{document}

\maketitle

\begin{abstract} \normalsize

\end{abstract}

The low-temperature series for the free energy density, pressure, magnetization and susceptibility of cubic ideal ferromagnets in weak
external magnetic fields are discussed within the effective Lagrangian framework up to three loops. The structure of the simple,
body-centered, and face-centered cubic lattice is taken into account explicitly. The expansion involves integer and half-integer powers of
the temperature. The corresponding coefficients depend on the magnetic field and on low-energy effective constants that can be expressed
in terms of microscopic quantities. Our formulas may also serve as efficiency or consistency check for other techniques like Green's
function methods, where spurious terms in the low-temperature expansion have appeared. We explore the sign and magnitude of the spin-wave
interaction in the pressure, magnetization and susceptibility, and emphasize that our effective field theory approach is fully systematic
and rigorous.


\maketitle

\section{Introduction}
\label{Intro}

The low-temperature expansion of the partition function for the three-dimensional ideal ferromagnet in a weak magnetic field has been
derived in Ref.~\citep{Hof11a} up to three-loop order within the systematic effective Lagrangian framework. The main intention of that
article was to go beyond Dyson's analysis, Refs.~\citep{Dys56a,Dys56b}, by evaluating the next-to-leading term in the spontaneous
magnetization caused by the spin-wave interaction. It was shown that the spin-wave interaction in the spontaneous magnetization -- beyond
the famous Dyson term of order $T^4$ -- already shows up at order $T^{9/2}$. In effective field theory language, Dyson's analysis
corresponds to two-loop order, while the next-to-leading contribution represents a three-loop effect.

However, in Ref.~\citep{Hof11a}, the three-dimensional ideal ferromagnet was assumed to be space rotation invariant. Here we abandon this
assumption and provide explicit expressions for the three types of cubic lattices for various thermodynamic quantities up to three-loop
order, including the free energy density, magnetization and susceptibility. We then explore how the spin-wave interaction in cubic ideal
ferromagnets manifests itself in the pressure, magnetization and susceptibility at low temperatures and in weak magnetic fields. While the
interaction is repulsive in the pressure, regarding the magnetization (susceptibility) the interaction contribution is negative (positive)
in the entire parameter regime where the effective analysis applies. We point out that the subleading term of order $T^{9/2}$ in the
magnetization enhances the Dyson term of order $T^4$, but it turns out to be very small.

The manifestation of the spin-wave interaction in the low-temperature behavior of three-dimensional ideal ferromagnets has been addressed
in numerous publications with various microscopic methods, pioneered by Dyson \citep{Dys56a,Dys56b} and Zittartz \citep{Zit65}. For recent
discussions of conceptual problems related to the simple cubic ideal ferromagnet in zero magnetic field, see Refs.~\citep{RPPK13,Rad15}.
Other relevant studies include Refs.~\citep{Mor58,Ogu60,Zub60,KL61,Sza61,TH62b,Sza63,SMS63,TC64,WJ64,Ort64,MT65a, MT65b,LD66,TMM66,CH67,
Wal67, VLP68,CH70,Coo70,CF73,Sza74,FC74, YW75,RL79,Bal80,Lol87,CH90,LWCY95,LC97,ICY02,CG12,CGS14,CGS15}. We stress that here we follow an
alternative route: the method of effective Lagrangians. As is well-known, the basic degrees of freedom at low temperatures in a
ferromagnet are the spin waves or magnons. The effective theory makes use of the fact that these are the Goldstone bosons resulting from
the spontaneously broken spin rotation symmetry. Systematic effective field theories for systems exhibiting spontaneous symmetry breaking
have been constructed a long time ago in particle physics \citep{Wei79,GL85,Leu94b}. The formalism has been transferred to the condensed
matter domain in Refs.~\citep{Leu94a,ABHV14}, showing that the effective Lagrangian method represents a rigorous and systematic tool to
investigate systems with a spontaneously broken symmetry in general. As we briefly review below, the method is based on a symmetry
analysis of the underlying system -- in the present study this is the microscopic Heisenberg ferromagnet.

In our final formulas for the thermodynamic observables, all low-energy constants of the effective theory have been expressed by
microscopic quantities like spin quantum number, exchange integral, and geometry factors. This matching between effective field theory and
microscopic theory is straightforward up to two-loop order, because Dyson has provided the respective coefficients in the free energy
density in his microscopic theory. However, in order to express all contributions arising at the three-loop level in terms of microscopic
constants, we have to evaluate the coefficient of a $T^{11/2}$-term in the free energy density that was not relevant in Dyson's analysis.
Fortunately, this coefficient is related to noninteracting magnons and thus poses no calculational problems. Remarkably, the three-loop
interaction contribution -- which is also of order $T^{11/2}$ -- only depends on the two leading-order effective constants, the spontaneous
magnetization at zero temperature and the spin stiffness, which already have been expressed in terms of microscopic quantities.

Conceptually, we observe that the numerical values of the various low-energy effective constants follow a hierarchical pattern: low-energy
constants that originate from higher-order pieces of the effective Lagrangian are gradually suppressed -- the effective field theory
approach is thus fully consistent.

Overall, we are able to give complete expressions for thermodynamic quantities up to three-loop order that involve microscopic constants
only and that explicitly take the structure of the three cubic crystal geometries into account. No approximations have been made in our
approach. While it is straightforward to pursue Dyson's program to higher orders in the low-temperature expansion using effective field
theory, it should be noted that within a microscopic framework -- Dyson's original approach, spin-wave theory, Schwinger-boson mean field
theory, Green's function theory, and yet other approaches -- such an endeavor would be formidable.

While the effective Lagrangian method is well-established in particle physics, in the condensed matter community the method is still not
very well known and not fully appreciated. In order to convince the community that we are dealing with a method that indeed deserves
attention, we like to enumerate some references where condensed matter problems have been addressed and solved within the effective
Lagrangian framework. Systems whose low-energy properties are governed by magnons include ferromagnetic spin chains \citep{GHKW10,Hof13a,
Hof14b}, as well as ferromagnets in two \citep{Hof12a,Hof12b,Hof14c} and three \citep{Hof11a,RPPK13,Rad15,Hof99a,RS99a,RS99b,Hof01,Hof02,
Hof11c,Hof15X} spatial dimensions. Antiferromagnetic and XY-type systems in two \citep{HN91,HN93,Hof10,Hof11b,Hof14d,Hof15} or three
\citep{HL90,Hof99b,RS00} spatial dimensions can be analyzed along the same lines. More complicated applications refer to the
antiferromagnetic precursors of high-temperature superconductors, which involve holes or electrons as additional degrees of freedom.
Systematic analyses were performed both for square \citep{KMW05,BKMPW06,BKPW06,BHKPW07,BHKMPW07,BHKPW08,JKHW09,BHKMPW09,VHJW12} and
honeycomb \citep{KBWHJW12, JKBWHW12,VHJW15} lattice geometries. Many of these studies, much like the present one, demonstrate that the
effective Lagrangian method is in fact superior to conventional condensed matter methods, as the analysis can be systematically taken to
higher orders in the perturbative expansion. We also point out that high-accuracy numerical simulations underline the correctness of the
effective Lagrangian approach \citep{WJ94,GHJNW09,JW11,Jia11,GHJPSW11}.

The rest of the paper is organized as follows. In Sec.~\ref{PartitionFunction} we discuss essential aspects of the effective Lagrangian
formalism and the evaluation of the partition function. The combination of Dyson's analysis with the effective field theory approach is
considered in Sec.~\ref{DysonEFT}, where the low-temperature series for the free energy density of cubic ideal ferromagnets are derived.
The behavior of the pressure, magnetization and susceptibility at low temperatures and weak magnetic fields is then discussed in
Sec.~\ref{PreMagSus}. We are interested in the manifestation of the spin-wave interaction, in particular, in the sign and magnitude of the
temperature-dependent interaction contributions. Finally, Sec.~\ref{Conclusions} contains our conclusions. Additional low-temperature
series for the energy density, heat capacity, and entropy density are provided in Appendix \ref{appendixA}.

\section{Partition Function up to Three Loops within the Effective Lagrangian Formalism}
\label{PartitionFunction}

In order not to be repetitive, here we only give a brief sketch of the effective Lagrangian method at finite temperature. For more details
we refer the reader to appendix A of Ref.~\citep{Hof11a} and the references mentioned there. Introductions to the effective Lagrangian
method are also provided by the pedagogic references \citep{Brau10,Bur07,Goi04,Sch03,Leu95}.

The fundamental degrees of freedom in the effective theory of magnetic systems are the spin waves or magnons. In the case of the
ferromagnet, there are two real components of the magnon field, which we denote by $U^a (a=1,2)$, and which are the first two components
of the three-dimensional magnetization unit vector $U^i = (U^a, U^3)$. In the effective Lagrangian, we then have time and space
derivatives that act on these fields. Terms with few derivatives dominate the low-energy behavior of the system, whereas terms with
higher-order derivatives are suppressed. The point is that this derivative expansion can be done systematically and that to a given order
-- in the present study we go up to three-loop order in the partition function -- only a finite number of terms and Feynman diagrams is
relevant. The effective Lagrangian needed for the present calculation is
\begin{eqnarray}
\label{Lagrangian}
{\cal L}_{eff} & = & \Sigma \frac{\epsilon_{ab} {\partial}_0 U^a U^b}{1+ U^3}
+ \Sigma \mu H U^3 - \mbox{$\frac{1}{2}$} F^2 {\partial}_r U^i {\partial}_r U^i + l_1 {( {\partial}_r U^i {\partial}_r U^i )}^2
\nonumber \\
& & + l_2 {( {\partial}_r U^i {\partial}_s U^i )}^2 + l_3 \Delta U^i \Delta U^i + c_1 U^i {\Delta}^3 U^i + d_1 U^i {\Delta}^4 U^i \, .
\end{eqnarray}
The first three terms represent the leading piece of the effective Lagrangian, ${\cal L}^2_{eff}$, that counts as order $p^2$. The
individual contributions involve one time derivative ($\partial_0$), two space derivatives ($\partial_r \partial_r$), and the magnetic
field that points into the third direction: ${\vec H}=(0,0,H)$. Note that we are dealing with nonrelativistic magnons displaying a
quadratic dispersion relation. Each term in the effective Lagrangian comes with an {\it a priori} unknown low-energy constant. In
${\cal L}^2_{eff}$ these are the spontaneous magnetization at zero temperature $\Sigma$, and the constant $F$ which is related to the
helicity modulus or spin stiffness $\gamma$ through $\gamma = F^2/\Sigma$. While the next-to-leading piece ${\cal L}^4_{eff}$ (order $p^4$)
in the effective Lagrangian involves three additional low-energy coupling constants ($l_1, l_2$ and $l_3$), the pieces ${\cal L}^6_{eff}$
(order $p^6$) and ${\cal L}^8_{eff}$ (order $p^8$) furthermore involve $c_1$ and $d_1$. The quantity $\Delta$ is the Laplace operator in
three dimensions.

It is important to point out that the structure of the above terms in the effective Lagrangian is a consequence of the symmetries of the
underlying microscopic theory. In the present case, the ideal\footnote{Following Dyson \citep{Dys56a}, {\it ideal} means that the exchange
couplings between nearest neighbors of the cubic lattice are purely isotropic.} Heisenberg ferromagnet ($J > 0$) in an external magnetic
field,
\label{HeisenbergModel}
\begin{equation}
{\cal H}_0 = -J \sum_{n.n.} {\vec S}_m \cdot {\vec S}_n - \mu \sum_n {\vec S}_n \cdot {\vec H} \, , \qquad J=const. \, , 
\end{equation}
exhibits spin rotation symmetry O(3), parity, time reversal, and invariance under the discrete symmetries of the cubic lattice. The
low-energy effective constants, on the other hand, are not fixed by symmetry. In Sec.~\ref{DysonEFT} we will determine and express all of
them through microscopic parameters by matching our effective results with Dyson's results, and by evaluating a specific higher-order term
explicitly within the microscopic framework. Our final expressions for the various thermodynamic quantities hence no longer contain
unknown low-energy coupling constants, but are explicit functions of the spin quantum number $S$, exchange integral $J$, and geometry
factors.

\begin{figure}
\begin{center}
\includegraphics[width=15cm]{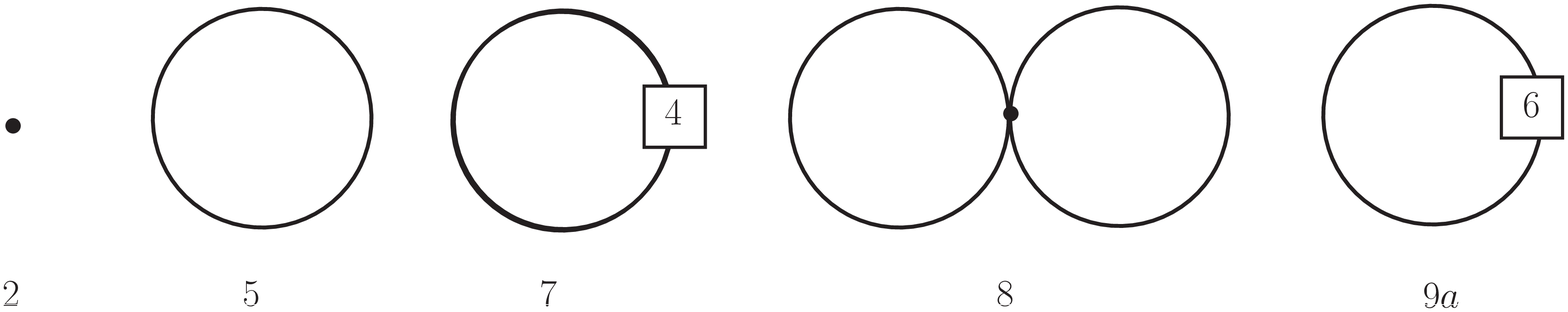}

\vspace{5mm}

\includegraphics[width=15cm]{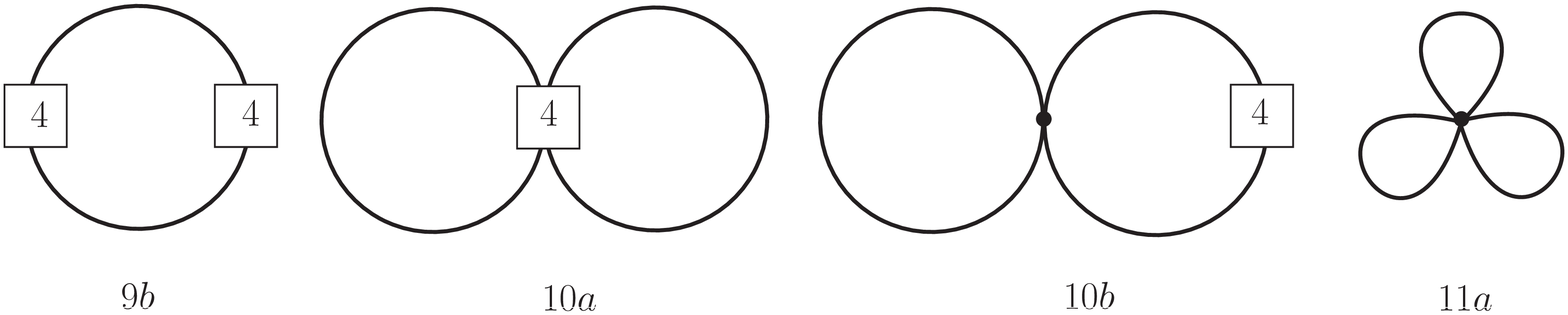}

\vspace{5mm}

\includegraphics[width=15cm]{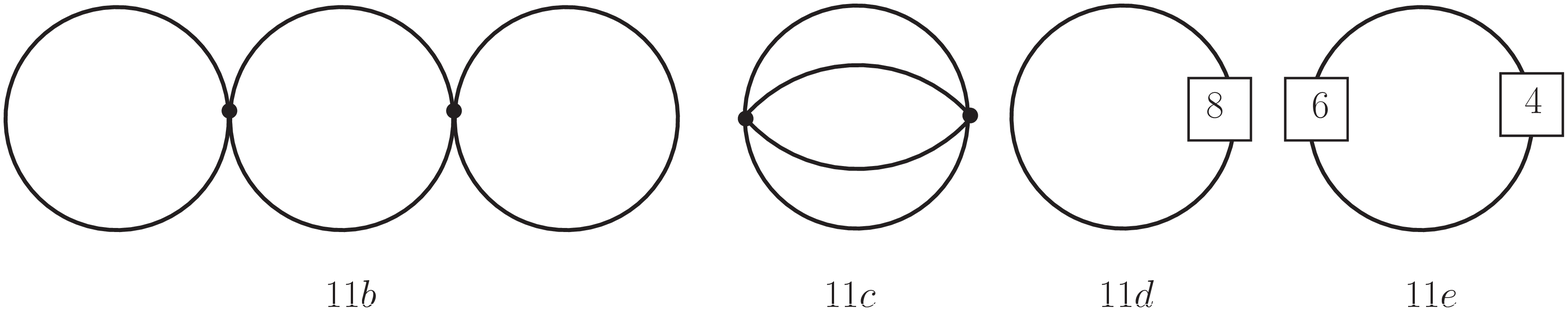}
\end{center}
\caption{Feynman graphs for the partition function of the three-dimensional ideal ferromagnet.}
\label{figure1}
\end{figure}

The perturbative evaluation of the partition function for the three-dimensional ideal ferromagnet has been presented in detail in
Ref.~\citep{Hof11a}. The relevant Feynman diagrams are shown in Fig.~\ref{figure1}. In the systematic effective expansion, loops are
suppressed by some power of momentum, energy, and temperature. In the context of ferromagnets, each loop in a Feynman diagram comes along
with a factor $p^{d_s}$ of momentum, where $d_s$ is the spatial dimension. In the present case, loops are thus suppressed by three powers
of momentum. Since we are dealing with nonrelativistic kinematics ($E \propto p^2$), each loop corresponds to a temperature power
$T^{3/2}$. The leading term in the free energy density is the Bloch term of order $T^{5/2} \propto p^5$ \citep{Blo30}, associated with the
one-loop graph 5. The two-loop diagram 8 (the three-loop diagrams 11a-c) are of order $T^4 \propto p^8$ ($T^{11/2} \propto p^{11}$),
because one (two) more loops are involved.

\begin{figure}
\label{figure2}
\begin{center}
\includegraphics[width=11cm]{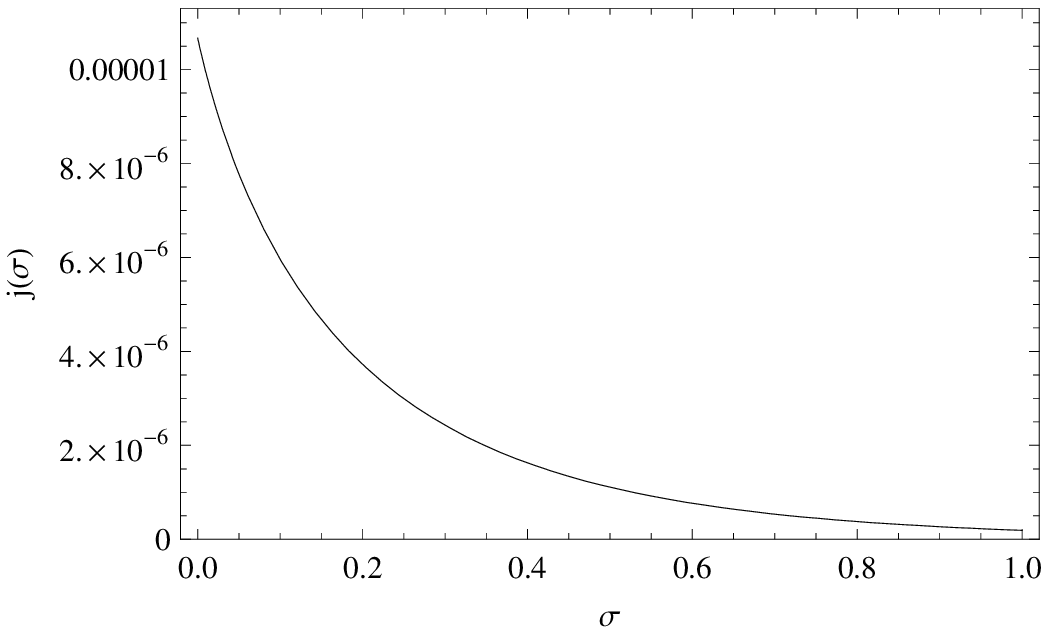}
\includegraphics[width=11cm]{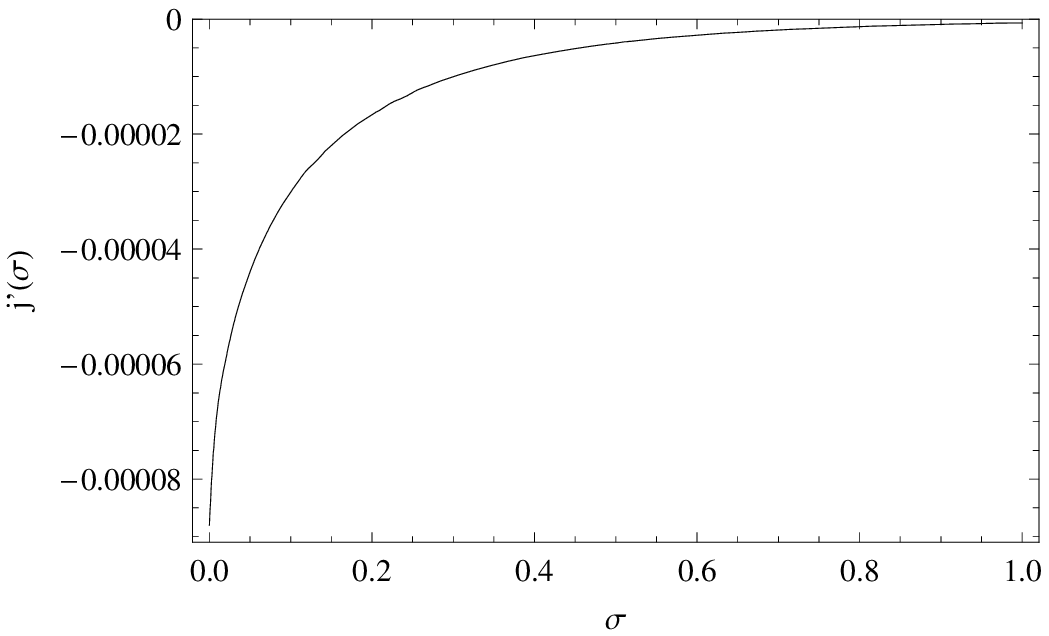}
\includegraphics[width=11cm]{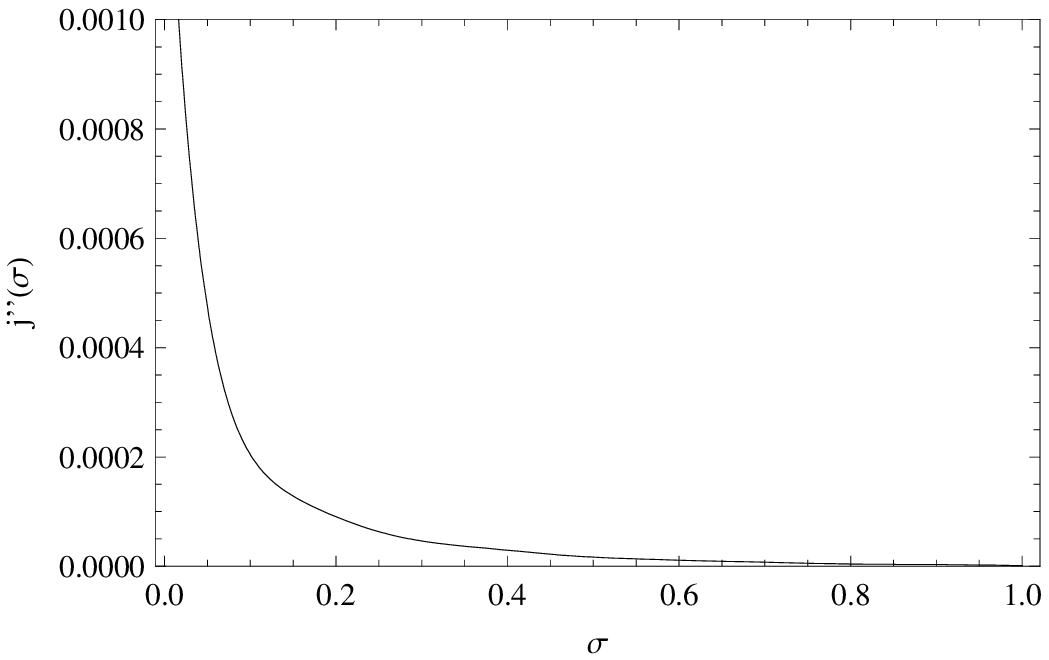}
\end{center}
\caption{The function $j(\sigma)$, as well as the derivatives $\frac{\partial j(\sigma)}{\partial \sigma}$ and
$\frac{\partial^2 j(\sigma)}{\partial \sigma^2}$, represent the interaction contribution at the three-loop level. The dimensionless
parameter $\sigma$ is defined by $\sigma = \mu H / T$.}
\end{figure}

Dyson's analysis that comprises all terms up to order $T^5$ in the free energy density, hence corresponds to a two-loop analysis in the
effective perspective that includes all diagrams up to order $p^{10}$ in Fig.~\ref{figure1}. Going one order beyond Dyson then means
including five additional diagrams -- diagrams 11a-e -- and working out the effective analysis up to three-loop order. The final result
for the low-temperature expansion of the free energy density in a weak magnetic field\footnote{In the next section we explain what is
meant by {\it low} temperature and {\it weak} magnetic field.} turns out to be \citep{Hof11a}
\begin{eqnarray}
\label{FreeCollectOrder11}
z(T,H) & = & - \Sigma \mu H - \frac{1}{8 {\pi}^{\frac{3}{2}} {\gamma}^{\frac{3}{2}}} \ T^{\frac{5}{2}} \sum^{\infty}_{n=1}
\frac{e^{- \sigma n}}{n^{\frac{5}{2}}}
- \frac{15 \,l_3}{16 {\pi}^{\frac{3}{2}} \Sigma {\gamma}^{\frac{7}{2}}} \ T^{\frac{7}{2}} \sum^{\infty}_{n=1}
\frac{e^{- \sigma n}}{n^{\frac{7}{2}}} \nonumber \\
& & - \frac{105}{32 {\pi}^{\frac{3}{2}} \Sigma {\gamma}^{\frac{9}{2}}} \, \Bigg( \frac{9 l^2_3}{2 \Sigma \gamma} - c_1 \Bigg) \ T^{\frac{9}{2}}
\sum^{\infty}_{n=1} \frac{e^{- \sigma n}}{n^{\frac{9}{2}}} \nonumber \\
& & - \frac{3 (8 l_1 + 6 l_2 + 5 l_3)}{128 {\pi}^3 {\Sigma}^2 {\gamma}^5} \ T^5 \, {\Bigg\{ \sum^{\infty}_{n=1}
\frac{e^{- \sigma n}}{n^{\frac{5}{2}}} \Bigg\}}^2 \nonumber \\
& & - \frac{945}{64 {\pi}^{\frac{3}{2}} \Sigma {\gamma}^{\frac{11}{2}}} \Bigg( d_1 - \frac{11 l_3 c_1}{\Sigma \gamma} \Bigg)
\, T^{\frac{11}{2}} \sum^{\infty}_{n=1} \frac{e^{- \sigma n}}{n^{\frac{11}{2}}} \nonumber \\
& & - \frac{1}{2 \Sigma^2 \gamma^{\frac{9}{2}}} \, j(\sigma) \, T^{\frac{11}{2}} + {\cal O}(T^6) \, ,
\end{eqnarray}
where the parameter $\sigma$ denotes the dimensionless ratio between magnetic field and temperature,
\begin{equation}
\sigma = \frac{\mu H}{T} \, .
\end{equation}
The quantity $j(\sigma)$ is a dimensionless function associated with the three-loop diagram 11c -- a plot of this function, as well as of
its first and second derivative that will be relevant in the magnetization and susceptibility, is provided in Fig.~\ref{figure2}.

The above series for the free energy density is the basic formula that we have derived in Ref.~\citep{Hof11a}. It is complete up to
three-loop order $p^{11} \propto T^{11/2}$. Still, it refers to a spatially isotropic three-dimensional ferromagnet. Furthermore,
Eq.~(\ref{FreeCollectOrder11}) is not very practical for the condensed matter community, as it features unknown low-energy effective
constants. In the following section, we will determine the actual values of these low-energy constants, and also fully take into account
the discrete symmetries of the simple cubic, body-centered cubic, and face-centered cubic lattice. 

\section{Dyson Theory and Effective Theory: Matching}
\label{DysonEFT}

Since the effective Lagrangian, Eq.~(\ref{Lagrangian}), is space rotation invariant, the reader may have severe doubts on how lattice
anisotropies should be accounted for in our effective description. First of all, it is well-known that lattice anisotropies only start
manifesting themselves at order ${\cal L}^4_{eff}$ \citep{HN93} -- the leading-order effective Lagrangian ${\cal L}^2_{eff}$ is strictly
space rotation invariant. This simply is an accidental symmetry that emerges at leading order which is not shared by higher-order pieces
of the effective Lagrangian.

On the other hand, assuming ${\cal L}^4_{eff}, {\cal L}^6_{eff}$, and ${\cal L}^8_{eff}$ to be space rotation invariant, is an idealization
that has to abandoned if one wants to apply our effective field theory predictions to actual condensed matter systems. Now Dyson has
provided complete expressions for all three types of cubic lattices up to order $p^{10} \propto T^5$ in the free energy density. So up to
two-loop order, all microscopic expressions are known and the effective field theory analysis merely served to corroborate Dyson's result
-- in particular, the structure of the temperature powers could be derived in an elegant and transparent manner. Had we included all
additional terms that arise in higher-order pieces of the effective Lagrangian due to lattice anisotropies, e.g., in ${\cal L}^4_{eff}$,
\begin{equation}
\sum_{r=1}^3 {\partial}_r {\partial}_r U^i {\partial}_r {\partial}_r U^i \, , \qquad
\sum_{r=1}^3 {\partial}_r U^i {\partial}_r U^i {\partial}_r U^k {\partial}_r U^k \, , \qquad \dots \, , 
\end{equation}
the structure of the low-temperature series for the free energy density would have been the same -- only the coefficients of the various
$T^m$-terms ($m=\mbox{$\frac{7}{2}$},\mbox{$\frac{9}{2}$},5,\mbox{$\frac{11}{2}$}$) would have involved additional low-energy effective
constants. While rigorous from a conceptual point of view, from a practical perspective it is unnecessary.

However, at the three-loop level $p^{11} \propto T^{11/2}$, we only have the effective formulas at hand -- a full microscopic analysis has
never been attempted. But we are in a fortunate situation. Inspecting the Feynman diagrams of order $p^{11}$ (see Fig.~\ref{figure1}), one
notices that the next-to leading pieces ${\cal L}^4_{eff}, {\cal L}^6_{eff}$, and ${\cal L}^8_{eff}$ only appear in one-loop graphs. Now
these graphs are easy to evaluate within the microscopic framework as we demonstrate below. Fortunately, the complicated three-loop graphs
11a-c only involve the leading-order effective Lagrangian ${\cal L}^2_{eff}$ which is strictly space rotation invariant and whose low-energy
constants $\Sigma$ and $F$ can easily be expressed in terms of microscopic quantities. In conclusion, by combining the microscopic
calculation (one-loop graphs 11d,e) and the effective calculation (three-loop graphs 11a-c), we can provide expressions for the free
energy density that no longer involve undetermined effective constants, but microscopic quantities only. Moreover the cubic lattice
geometry is fully accounted for.

Let us first consider the matching procedure up to two-loop order $p^{10} \propto T^5$, by comparing the effective low-temperature series
for the free energy density, Eq.~(\ref{FreeCollectOrder11}), with the microscopic series derived by Dyson, Eq.~(131) in
Ref.~\citep{Dys56b}. This allows us to extract the actual values of the low-energy effective constants contained in ${\cal L}^2_{eff},
{\cal L}^4_{eff}$, and ${\cal L}^6_{eff}$ as follows:
\begin{eqnarray}
\label{matchingTwoLoop}
\Sigma & = & \Big\{ 1, 2, 4 \Big\} \, \frac{S}{a^3} \, , \nonumber \\
\gamma & = & 2 S J a^2 \, , \nonumber \\
F^2 & = & \gamma \Sigma = \Big\{ 1, 2, 4 \Big\} \, \frac{2 S^2 J}{a} \, , \nonumber \\
l_3 & = & \Bigg\{ \frac{1}{20}, \frac{3}{40}, \frac{1}{10} \Bigg\} \, S^2 J a \, , \nonumber \\
\mbox{$\frac{8}{5}$} l_1 + \mbox{$\frac{6}{5}$} l_2 & \approx & \Bigg\{ \frac{-0.00861 + 0.153 S}{(10 S - 1)S},
\frac{-0.00983 + 0.257 S}{(16 S - 1)S}, \nonumber \\
& & \frac{-0.0115 + 0.409 S}{(24 S - 1)S} \Bigg\} \, S^2 J a \, , \nonumber \\
c_1 & = & \Bigg\{ \frac{1}{1400}, \frac{37}{33600}, \frac{13}{11200} \Bigg\} \, S^2 J a^3 \, .
\end{eqnarray}
The various microscopic quantities are spin quantum number ($S$), exchange integral ($J$), and lattice constant ($a$). The three values in
the brace associated with $\Sigma, F^2, l_3,\mbox{$\frac{8}{5}$} l_1 + \mbox{$\frac{6}{5}$} l_2 $, and $c_1$, refer to the simple cubic,
body-centered cubic, and face-centered cubic lattice, respectively.

The effective constant $\gamma=F^2/\Sigma$ , that appears in the leading term of the ferromagnetic dispersion relation,
\begin{equation}
\omega = \gamma {\vec k}^2 \, ,
\end{equation}
is universal -- it is not affected by lattice anisotropies and hence is the same for all three cubic crystals. On the other hand, the
spontaneous magnetization $\Sigma$, the constant $F^2$, as well as the higher-order effective constants $l_1, l_2, l_3, c_1$ -- related
${\cal L}^4_{eff}$ and ${\cal L}^6_{eff}$ -- depend on the specific cubic lattice. In particular, $\Sigma$ measures the spin per unit
volume $a^3$. Since the primitive cells of the $bcc$ and $f\!cc$ lattices (that each contain one atom or one spin) have volumes
$\mbox{$\frac{1}{2}$} a^3$ and $\mbox{$\frac{1}{4}$} a^3$, respectively, appropriate scaling factors appear in the effective constant
$\Sigma$. Note that the effective constant $l_3$ (unlike $l_1$ and $l_2$) already appears in the $T^{7/2}$-coefficient in the free energy
density (originating from the one-loop graph 7), such that it can be extracted from there.

Furthermore, for the combination $\mbox{$\frac{8}{5}$} l_1 + \mbox{$\frac{6}{5}$} l_2$ of effective constants, analytic expressions
related to Watson integrals \citep{Wat39} can be provided for all three cubic lattices,
\begin{eqnarray}
\label{matchingTwoLoopWatson}
\mbox{$\frac{8}{5}$} l_1 + \mbox{$\frac{6}{5}$} l_2 & = & \Bigg\{ \frac{-I_{sc} + 4 S + 10 S I_{sc}}{60 (10 S - 1)S},
\frac{-I_{bcc} + 4 S + 16 S I_{bcc}}{40(16 S - 1)S}, \nonumber \\
& & \frac{-I_{f\!cc} + 4 S + 24 S I_{f\!cc}}{30(24 S - 1)S} \Bigg\} \, S^2 J a \, .
\end{eqnarray}
Using the approximate numerical values for the quantities $I_{\alpha}$,
\begin{equation}
I_{sc} = 0.5164 \, , \qquad I_{bcc} = 0.3932 \, , \qquad I_{f\!cc} = 0.3447 \, ,
\end{equation}
the expressions for $\mbox{$\frac{8}{5}$} l_1 + \mbox{$\frac{6}{5}$} l_2$ given in Eq.~(\ref{matchingTwoLoop}) are obtained.

We now proceed with the analysis at the three-loop level $p^{11} \propto T^{11/2}$. We first discuss the one-loop graphs 11d and 11e. These
graphs are easy to evaluate in the microscopic theory, as they correspond to noninteracting magnons. We follow the method advocated in the
original articles by Dyson \citep{Dys56a,Dys56b}. The basic formula for the free energy density at finite temperature is
\begin{equation}
\label{zFree}
z_{free} = \frac{T}{8 \pi^3} \, \int d^3k \ln \Big( 1- e^{-\omega({\vec k})/T} \Big) \, , \qquad {\vec k} = (k_1, k_2, k_3) \, .
\end{equation}
The dispersion relation $\omega({\vec k})$ is obtained from the structure factor $\gamma(\vec k)$,
\begin{equation}
\gamma(\vec k) = \sum_{n.n.} \exp\Big({i {\vec k} \cdot {\vec \delta}}\Big) \, ,
\end{equation}
where the vectors $\vec \delta$ connect a given lattice site with its nearest neighbors. For the three types of cubic lattices one arrives
at
\begin{eqnarray}
\gamma_{sc}(\vec k)  & = & 2 \cos(k_1 a) + 2 \cos(k_2 a) + 2 \cos(k_3 a) \, , \nonumber \\
\gamma_{bcc}(\vec k) & = & 8 \cos(\frac{k_1 a}{2}) \cos(\frac{k_2 a}{2}) \cos(\frac{k_3 a}{2}) \, , \\
\gamma_{f\!cc}(\vec k) & = & 4 \cos(\frac{k_1 a}{2}) \cos(\frac{k_2 a}{2}) + 4 \cos(\frac{k_1 a}{2}) \cos(\frac{k_3 a}{2})
+ 4 \cos(\frac{k_2 a}{2}) \cos(\frac{k_3 a}{2}) \, . \nonumber
\end{eqnarray}
Expanding these expressions in the momenta then leads to
\begin{eqnarray}
\label{kseries}
\omega(\vec k) & = & 2 S J \, \Big( z_{\alpha} - \gamma(\vec k) \Big) + \mu H \nonumber \\
& = & \mu H + 2 S J (k_1^2 + k_2^2 + k_3^2) a^2 + \gamma_{1A} S J (k_1^4 + k_2^4 + k_3^4) a^4 \nonumber \\
& & + \gamma_{1B} S J ( k_1^2 k_2^2 + k_1^2 k_3^2 + k_2^2 k_3^2) a^4
+ \gamma_{2A} S J (k_1^6 + k_2^6 + k_3^6) a^6  \nonumber \\
& & + \gamma_{2B} S J k_1^2 k_2^2 k_3^2 a^6 + \gamma_{2C} S J k_1^4 k_2^2 a^6 + \dots \, ,
\end{eqnarray}
where $z_{\alpha}$ is the number of nearest neighbors of a given lattice site,
\begin{equation}
z_{sc} = 6 \, , \qquad z_{bcc} = 8 \, , \qquad z_{f\!cc} = 12 \, .
\end{equation}
The leading term in the dispersion relation, as stated before, is identical for all three cubic lattices. The anisotropies manifest
themselves in all other terms whose coefficients hence depend on the specific type of lattice geometry. Using spherical coordinates to
perturbatively evaluate the integral (\ref{zFree}) with the expansion (\ref{kseries}), the final result for the total one-loop
contribution at order $T^{11/2}$ in the free energy density reads
\begin{equation}
\label{11Microscopic}
z_{11d + 11e} = - \frac{d_{\alpha}}{\sqrt{2} \pi^{3/2} J^{9/2} S^{9/2}} \,
\Bigg( \sum^{\infty}_{n=1} \frac{e^{- \mu H n /T}}{n^{\frac{11}{2}}} \Bigg) \, T^{11/2} \, .
\end{equation}
The respective coefficients $d_{\alpha}$ for the simple cubic, body-centered cubic, and face-centered cubic lattice are
\begin{equation}
\label{identificationD}
d_{sc} = \frac{281}{1048576} \, , \qquad d_{bcc} = \frac{6419}{134217728} \, , \qquad d_{f\!cc} = \frac{107}{16777216} \, .
\end{equation}
Note that, for the simple cubic lattice, a microscopic expression for the coefficient $d_{\alpha}$ has also been derived in
Ref.~\citep{RPPK13} -- it perfectly coincides with our result. Finally, by matching the microscopic formula, Eq.~(\ref{11Microscopic}),
with the effective field theory result, Eq.~(\ref{FreeCollectOrder11}), we can express the low-energy effective constant $d_1$,
originating from ${\cal L}^8_{eff}$, through microscopic quantities as
\begin{equation}
\label{matchingThreeLoop}
d_1 = \Bigg\{ \frac{1343}{1728000}, \frac{248387}{387072000}, \frac{9211}{24192000} \Bigg\} \, S^2 J a^5 \, .
\end{equation}

We are left with the other type of contributions also arising at order $T^{11/2}$ in the free energy density. These are related to the
three-loop interaction graphs 11a-c that exclusively involve vertices from the leading piece ${\cal L}^2_{eff}$. A microscopic expression
for the last term in Eq.~(\ref{FreeCollectOrder11}) can thus immediately be given, using the first two matching equations
(\ref{matchingTwoLoop}).

After these manipulations we can now provide the low-temperature series for the free energy per atom\footnote{Note that
from here on -- since we now provide microscopic expressions -- we follow Dyson's convention where "free energy density" corresponds to
free energy per atom (or spin).} of cubic ideal ferromagnets up to three-loop order, from which any other thermodynamic quantity may be
derived,
\begin{eqnarray}
\label{fed}
z(T,H) & = & - \mu H S - \frac{b_{\alpha}}{16 \sqrt{2} \pi^{3/2} J^{3/2} S^{3/2}} \, \Bigg( \sum^{\infty}_{n=1}
\frac{e^{- \mu H n/T}}{n^{\frac{5}{2}}} \Bigg) \ T^{\frac{5}{2}} \nonumber \\
& & - \frac{l_{\alpha}}{\sqrt{2} \pi^{3/2} J^{5/2} S^{5/2}} \, \Bigg( \sum^{\infty}_{n=1}
\frac{e^{- \mu H n/T}}{n^{\frac{7}{2}}} \Bigg) \ T^{\frac{7}{2}}
- \frac{c_{\alpha}}{\sqrt{2} \pi^{3/2} J^{7/2} S^{7/2}} \,
\Bigg( \sum^{\infty}_{n=1} \frac{e^{- \mu H n/T}}{n^{\frac{9}{2}}} \Bigg) \ T^{\frac{9}{2}} \nonumber \\
& & - \frac{L_{\alpha}}{\pi^3 J^4 S^5} \, 
{\Bigg( \sum^{\infty}_{n=1} \frac{e^{- \mu H n/T}}{n^{\frac{5}{2}}} \Bigg)}^2 T^5
- \frac{ d_{\alpha}}{\sqrt{2} \pi^{3/2} J^{9/2} S^{9/2}} \,
\Bigg( \sum^{\infty}_{n=1} \frac{e^{- \mu H n/T}}{n^{\frac{11}{2}}} \Bigg) \, T^{\frac{11}{2}} \nonumber \\
& & - \, \frac{j_{\alpha}}{32 \sqrt{2} J^{9/2} S^{13/2}} \, j(\mu H/T) \, T^{\frac{11}{2}} + {\cal O}(T^6) \, .
\end{eqnarray}
The coefficients $b_{\alpha}, l_{\alpha}, L_{\alpha}, c_{\alpha}, d_{\alpha}, j_{\alpha}$ for the three cubic lattices are:
\begin{itemize}

\vspace{-3mm}

\item Simple cubic lattice
\begin{eqnarray}
\label{identificationSC}
b_{sc} & = & 1 \, , \nonumber \\
l_{sc} & = & \frac{3}{512} \approx 5.86 \times 10^{-3} \, , \nonumber \\
L_{sc} & \approx & \frac{0.00183 S^2 + 0.000376 S - 0.0000315}{(10 S-1)S} \approx 3.07 \times 10^{-4} \ (S=\mbox{$\frac{1}{2}$}) \, ,
\nonumber \\
c_{sc} & = & \frac{33}{32768} \approx 1.01 \times 10^{-3} \, , \nonumber \\
d_{sc} & = & \frac{281}{1048576} \approx 2.68 \times 10^{-4} \, , \nonumber \\
j_{sc} & = & 1 \, .
\end{eqnarray}

\end{itemize}

\begin{itemize}

\vspace{-3mm}

\item Body-centered cubic lattice
\begin{eqnarray}
\label{identificationBCC}
b_{bcc} & = & \frac{1}{2} \, , \nonumber \\
l_{bcc} & = & \frac{9}{4096} \approx 2.20 \times 10^{-3} \, , \nonumber \\
L_{bcc} & \approx & \frac{0.000549 S^2 + 0.0000834 S - 0.00000450}{(16 S-1)S} \approx 4.99 \times 10^{-5} \ (S=\mbox{$\frac{1}{2}$}) \, ,
\nonumber \\
c_{bcc} & = & \frac{281}{1048576} \approx 2.68 \times 10^{-4} \, , \nonumber \\
d_{bcc} & = & \frac{6419}{134217728} \approx 4.78 \times 10^{-5} \, , \nonumber \\
j_{bcc} & = &  \frac{1}{8} \, .
\end{eqnarray}

\end{itemize}

\begin{itemize}

\vspace{-3mm}

\item Face-centered cubic lattice
\begin{eqnarray}
\label{identificationFCC}
b_{f\!cc} & = & \frac{1}{4} \, , \nonumber \\
l_{f\!cc} & = & \frac{3}{4096} \approx 7.32 \times 10^{-4} \, , \nonumber \\
L_{f\!cc} & \approx & \frac{0.000137 S^2 + 0.0000177 S - 0.000000657}{(24 S-1)S} \approx 7.73 \times 10^{-6} \ (S=\mbox{$\frac{1}{2}$}) \, ,
\nonumber \\
c_{f\!cc} & = & \frac{15}{262144} \approx 5.72 \times 10^{-5} \, , \nonumber \\
d_{f\!cc} & = & \frac{107}{16777216} \approx 6.38 \times 10^{-6} \, , \nonumber \\
j_{f\!cc} & = & \frac{1}{64} \, .
\end{eqnarray}
\end{itemize}

It should be pointed out that all low-energy effective constants have been expressed through microscopic quantities. We also emphasize
that any future alternative derivation of the $T^{11/2}$-contributions in the free energy density of cubic ideal ferromagnets, must end up
with the result we have provided. In this perspective, our expression may serve as an efficiency or consistency check for other techniques
like Green's function methods or spin-wave theory, and may hence help to understand why some of these calculations sometimes go wrong. In
that context, an interesting and comprehensive analysis of spurious terms in the free energy density -- in particular, of the $T^4$-term
that haunted the literature for years -- has been given in Refs.~\citep{RPPK13,Rad15}.

\begin{figure}
\begin{center}
\includegraphics[width=12.0cm]{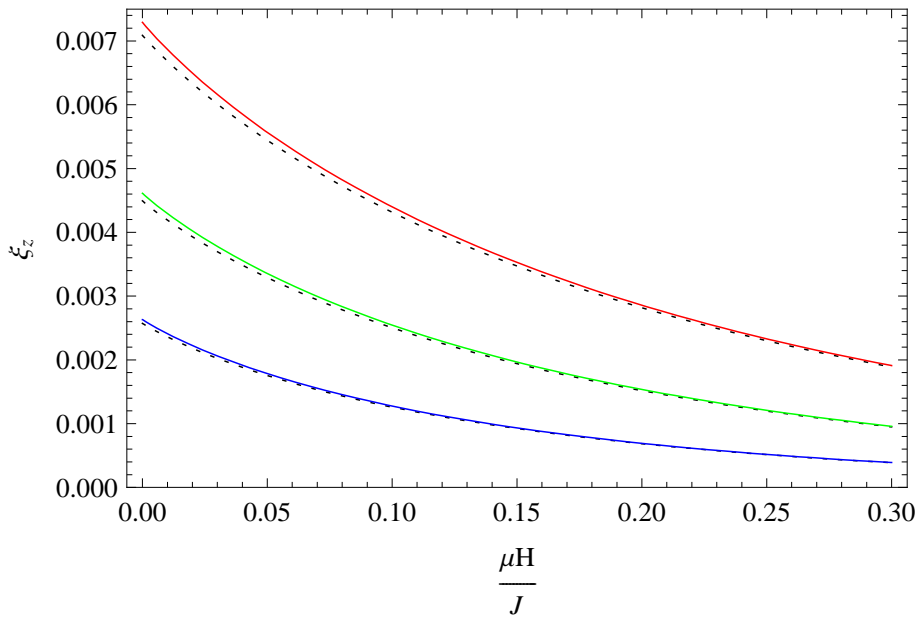}
\end{center}
\caption{[Color online] The three-loop ($T^{11/2}$) interaction term represents a small correction to the two-loop ($T^5$) interaction
contribution (dashed black curves) in the free energy density of the simple cubic ideal ferromagnet ($S=\mbox{$\frac{1}{2}$}$), according
to Eq.~(\ref{measureZ}). The temperatures are $T/J = \{ \mbox{$\frac{4}{10}$}, \mbox{$\frac{5}{10}$}, \mbox{$\frac{6}{10}$} \}$ from
bottom to top in the figure.}
\label{fig3}
\end{figure}

\begin{figure}
\begin{center}
\includegraphics[width=12.0cm]{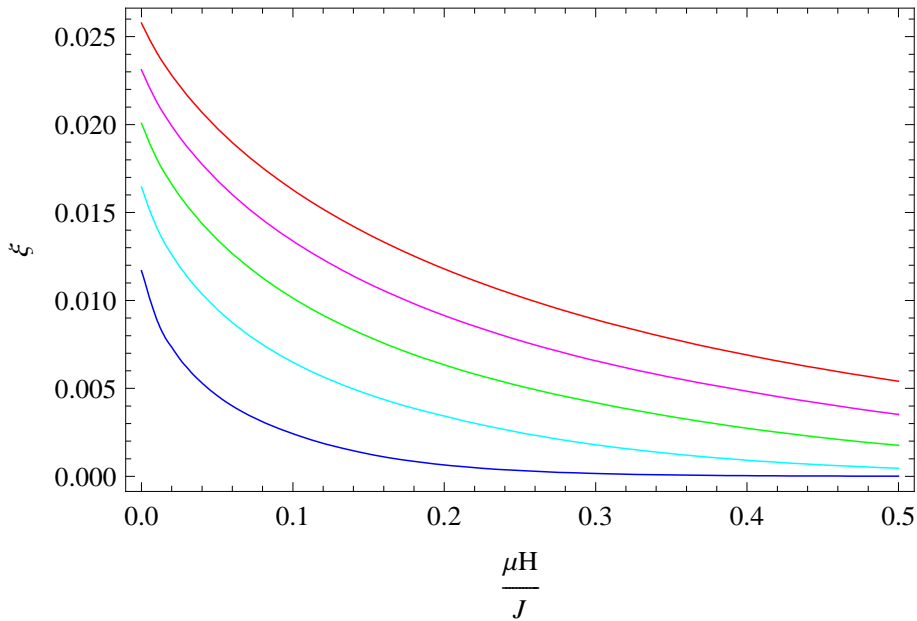}
\end{center}
\caption{[Color online] Relative strength of the three-loop ($T^{11/2}$) interaction contribution in the free energy density of the simple
cubic ideal ferromagnet ($S=\mbox{$\frac{1}{2}$}$) as a function of the magnetic field, according to Eq.~(\ref{measureR}). The
temperatures are $T/J = \{\mbox{$\frac{1}{10}$},\mbox{$\frac{2}{10}$},\mbox{$\frac{3}{10}$},\mbox{$\frac{4}{10}$},\mbox{$\frac{5}{10}$}\}$
from bottom to top in the figure.}
\label{fig4}
\end{figure}

In his pioneering work, Dyson concluded that the spin-wave interaction in cubic ideal ferromagnets is very weak. In order to have a
quantitative measure, we consider the dimensionless ratio
\begin{equation}
\label{measureZ}
\xi_z(T,H) = \frac{16 \sqrt{2} \pi^{3/2} J^{3/2} S^{3/2}}{T^{5/2}} \, z_{int}(T,H) \, ,
\end{equation}
that measures the strength and the sign of the interaction contributions in the free energy density with respect to the leading Bloch term
(proportional to $T^{5/2}$ in $z$). In Fig.~\ref{fig3}, we depict the two-loop interaction contribution (Dyson term), as well as the sum
of the two- and three-loop interaction contribution. The curves refer to the three temperatures $T/J = \{\mbox{$\frac{4}{10}$},
\mbox{$\frac{5}{10}$},\mbox{$\frac{6}{10}$}\}$ and to $S=\mbox{$\frac{1}{2}$}$. Indeed, the spin-wave interaction is very weak at low
temperatures and weak magnetic fields. While the interaction increases as the temperature rises, it decreases in stronger magnetic fields,
and is largest in the limit ${\vec H} \to 0$. Still, even for the temperature $T/J = \mbox{$\frac{6}{10}$}$, the interaction only makes up
about one percent relative to the Bloch term. The three-loop correction has the same sign as the two-loop contribution and thus enhances
the spin-wave interaction. However, as we further illustrate in Fig.~\ref{fig4}, the tree-loop correction is indeed very small. The ratio
\begin{equation}
\label{measureR}
\xi(T,H) = \frac{z^{3-loop}_{int}(T,H)}{z^{2-loop}_{int}(T,H) + z^{3-loop}_{int}(T,H)}
\end{equation}
is less than three percent in zero magnetic field at the temperature $T/J = \mbox{$\frac{5}{10}$}$ and $S=\mbox{$\frac{1}{2}$}$. Note that
the spin-wave interaction is most pronounced for the case $S=\mbox{$\frac{1}{2}$}$ we have depicted. Moreover, while the above plots refer
to the simple cubic lattice, qualitatively similar results emerge for the $bcc$ and $f\!cc$ lattices.

Before we discuss the manifestation of the spin-wave interaction in the pressure, magnetization and susceptibility, we would like to
comment on the consistency and domain of validity of the effective expansion.

The consistency concerns the order of magnitude of the effective constants and their relative suppression. This is illustrated in the
following compilation of low-energy effective constants that refers to $S=\mbox{$\frac{1}{2}$}$:
\begin{eqnarray}
\label{matchingLECNumeric}
\Sigma & = & \Big\{ \mbox{$\frac{1}{2}$}, 1, 2 \Big\} \, \frac{1}{a^3} \, , \nonumber \\
\gamma & = & J a^2 \, , \nonumber \\
F^2 & = & \gamma \Sigma = \Big\{ \mbox{$\frac{1}{2}$}, 1, 2 \Big\} \, \frac{J}{a} \, , \nonumber \\
l_3 & \approx & \Big\{ 1.25, 1.88, 2.50 \Big\} \times 10^{-2} \, J a \, , \nonumber \\
\mbox{$\frac{8}{5}$} l_1 + \mbox{$\frac{6}{5}$} l_2 & \approx & \Big\{ 3.39, 3.39, 3.51 \Big\} \times 10^{-2} \, J a \, , \nonumber \\
c_1 & \approx & \Big\{ 1.79, 2.75, 2.90 \Big\} \times 10^{-4} \, J a^3 \, , \nonumber \\
d_1 & \approx & \Big\{ 1.94, 1.60, 0.952 \Big\} \times 10^{-4} \, J a^5 \, .
\end{eqnarray}
Clearly, a hierarchical pattern in the numerical values of these effective constants emerges. Compared to the leading-order constants
$\Sigma, \gamma$ and $F^2$ (related to ${\cal L}^2_{eff}$), the next-to-leading order constant $l_3$ and the combination
$\mbox{$\frac{8}{5}$} l_1 + \mbox{$\frac{6}{5}$} l_2$ (from ${\cal L}^4_{eff}$), are about two orders of magnitude smaller. In turn, the
constant $c_1$ from ${\cal L}^6_{eff}$, and $d_1$ from ${\cal L}^8_{eff}$, are smaller than the couplings $l_i$. The hierarchy can further
be observed in the successive suppression of the effective constants by the scale $\Lambda$,
\begin{equation}
\Lambda \propto \frac{1}{a} \, ,
\end{equation}
as follows ($a$ is the lattice constant):
\begin{equation}
F^2 \propto \Lambda \, , \qquad l_i \propto \frac{1}{\Lambda} \, , \qquad c_1 \propto \frac{1}{\Lambda^3} \, , \qquad
d_1 \propto \frac{1}{\Lambda^5} \, . 
\end{equation}
This implies that the effective framework is consistent: in the derivative (or momentum) expansion, higher-order terms are less important
since they are suppressed by powers of $p/\Lambda$.

Finally, regarding the validity range of the effective expansion, we clarify what is meant by {\it low} temperature and {\it weak}
magnetic field. Both quantities must be small relative to the intrinsic scale of the underlying theory. In the present context, the
underlying theory is the microscopic Heisenberg Hamiltonian of the ferromagnet, where the relevant scale can be identified with the
exchange integral $J$. We may also invoke the Curie temperature $T_C$, at which the internal O(3) spin rotation symmetry is restored, and
the spin-wave picture breaks down. Since $T_C$ is of the order of $J$ \citep{Dys56a,Dys56b,RPPK13}, we conclude that, for the effective
expansion to be valid, temperature and magnetic field must be within the range \begin{equation}
T, \mu H \ \lesssim \ 0.5 \, J \, .
\end{equation}
Note that, while the temperature and the magnetic field must be small, their ratio $\sigma = \mu H /T$ can take any value.

\section{Manifestation of the Spin-Wave Interaction}
\label{PreMagSus}

In this section we discuss the manifestation of the spin-wave interaction in the pressure, magnetization and susceptibility at low
temperature and weak magnetic field. We first consider the pressure that is obtained from the free energy density by
\begin{equation}
P = z_0 - z \, .
\end{equation}
The low-temperature series amounts to
\begin{equation}
\label{Pressure}
P(T,H) = h_0 T^{\frac{5}{2}} + h_1 T^{\frac{7}{2}} + h_2 T^{\frac{9}{2}} + h_3 T^5 + h_4 T^{\frac{11}{2}} + {\cal O}(T^6) \, ,
\end{equation}
where the quantities $h_i$ depend on the ratio $\sigma=\mu H/T$ and take the form
\begin{eqnarray}
h_0 & = & \frac{b_{\alpha}}{16 \sqrt{2} \pi^{3/2} J^{3/2} S^{3/2}} \, \sum^{\infty}_{n=1}
\frac{e^{- \mu H n/T}}{n^{\frac{5}{2}}} \, , \nonumber \\
h_1 & = & \frac{l_{\alpha}}{\sqrt{2} \pi^{3/2} J^{5/2} S^{5/2}} \, \sum^{\infty}_{n=1}
\frac{e^{- \mu H n/T}}{n^{\frac{7}{2}}} \, , \nonumber \\
h_2 & = & \frac{c_{\alpha}}{\sqrt{2} \pi^{3/2} J^{7/2} S^{7/2}} \,
\sum^{\infty}_{n=1} \frac{e^{- \mu H n/T}}{n^{\frac{9}{2}}} \, , \nonumber \\
h_3 & = & \frac{L_{\alpha}}{\pi^3 J^4 S^5} \, 
{\Bigg\{ \sum^{\infty}_{n=1} \frac{e^{- \mu H n/T}}{n^{\frac{5}{2}}} \Bigg\}}^2 \, , \nonumber \\
h_4 & = & \frac{ d_{\alpha}}{\sqrt{2} \pi^{3/2} J^{9/2} S^{9/2}} \,
\sum^{\infty}_{n=1} \frac{e^{- \mu H n/T}}{n^{\frac{11}{2}}} \, 
+ \frac{j_{\alpha}}{32 \sqrt{2} J^{9/2} S^{13/2}} \, j(\mu H/T) \, .
\end{eqnarray}
The coefficients $b_{\alpha}, l_{\alpha}, L_{\alpha}, c_{\alpha}, d_{\alpha}$ and $j_{\alpha}$ are provided in Eq.~(\ref{identificationSC}),
Eq.~(\ref{identificationBCC}) and Eq.~(\ref{identificationFCC}) for the three types of cubic lattices. Note that the limit $H \to 0$ in
the low-temperature expansion for the pressure is well-defined. The infinite sums simply reduce to Riemann zeta functions, and the
dimensionless three-loop function at the origin takes the value $j(0) = 1.07 \times 10^{-5}$ \citep{Hof11a,RPPK13}.\footnote{We point out
that the numerical analysis used to determine $j(0)$ in Ref.~\citep{RPPK13} is based on lattice regularization, while the numerical
analysis we used in Ref.~\citep{Hof11a} is based on dimensional regularization. Both approaches lead to the same result.}

\begin{figure}
\begin{center}
\includegraphics[width=12.0cm]{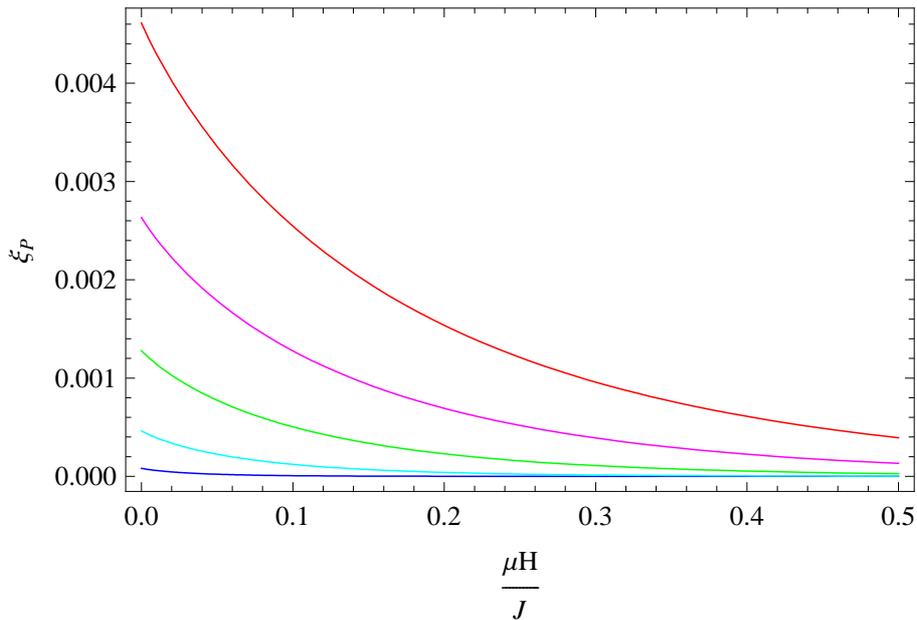}
\end{center}
\caption{[Color online] Sign and magnitude of the spin-wave interaction in the pressure (simple cubic ideal ferromagnet,
$S=\mbox{$\frac{1}{2}$}$) as a function of the magnetic field, according to Eq.~(\ref{xiP}). The curves refer to the temperatures
$T/J = \{ \mbox{$\frac{1}{10}$},\mbox{$\frac{2}{10}$}, \mbox{$\frac{3}{10}$},\mbox{$\frac{4}{10}$}, \mbox{$\frac{5}{10}$} \}$ from bottom
to top in the figure.}
\label{figure5}
\end{figure}

To capture sign and magnitude of the spin-wave interaction in the pressure -- relative to the leading Bloch term -- we define the
dimensionless ratio $\xi_P$, 
\begin{equation}
\label{xiP}
\xi_P(T,H) = \frac{16 \sqrt{2} \pi^{3/2} J^{3/2} S^{3/2}}{T^{5/2}} \, P_{int}(T,H) \, .
\end{equation}
In Fig.~\ref{figure5} we consider the total interaction contribution (sum of two-loop and three-loop term). The plots are for the simple
cubic lattice and $S=\mbox{$\frac{1}{2}$}$. The five curves refer to the temperatures $T/J = \{ \mbox{$\frac{1}{10}$},\mbox{$\frac{2}{10}$},
\mbox{$\frac{3}{10}$},\mbox{$\frac{4}{10}$},\mbox{$\frac{5}{10}$} \}$. The interaction is largest when the magnetic field is switched off,
and becomes stronger at more elevated temperatures. Note that the parameter $\xi_P(T,H)$ takes positive values: the spin-wave interaction
in the pressure hence is repulsive at low temperatures and weak magnetic fields. Here we are invoking the picture of the non-ideal magnon
gas, where a positive sign of the overall interaction contribution in the pressure signals repulsion.

We proceed with the low-temperature expansion for the magnetization,
\begin{equation}
M(T,H) = - \frac{\partial z(T,H)}{\partial(\mu H)} \, .
\end{equation}
Up to three-loop order $T^{9/2}$, we obtain
\begin{equation}
\frac{M(T,H)}{S} = 1 - a_0 T^{\frac{3}{2}} - a_1 T^{\frac{5}{2}} - a_2 T^{\frac{7}{2}} - a_3 T^4 - a_4 T^{\frac{9}{2}} + {\cal O}(T^5) \, .
\end{equation}
The coefficients $a_i$ depend on the ratio $\sigma=\mu H/T$ and are given by
\begin{eqnarray}
\label{MagnetCollectOrder9}
a_0 & = & \frac{b_{\alpha}}{16 \sqrt{2} \pi^{3/2} J^{3/2} S^{5/2}} \, \sum^{\infty}_{n=1}
\frac{e^{- \mu H n/T}}{n^{\frac{3}{2}}} \, , \nonumber \\
a_1 & = & \frac{l_{\alpha}}{\sqrt{2} \pi^{3/2} J^{5/2} S^{7/2}} \, \sum^{\infty}_{n=1}
\frac{e^{- \mu H n/T}}{n^{\frac{5}{2}}} \, , \nonumber \\
a_2 & = & \frac{c_{\alpha}}{\sqrt{2} \pi^{3/2} J^{7/2} S^{9/2}} \,
\sum^{\infty}_{n=1} \frac{e^{- \mu H n/T}}{n^{\frac{7}{2}}} \, , \nonumber \\
a_3 & = & \frac{2 L_{\alpha}}{\pi^3 J^4 S^6} \, 
\sum^{\infty}_{n=1} \frac{e^{- \mu H n/T}}{n^{\frac{3}{2}}} \, \sum^{\infty}_{m=1} \frac{e^{- \mu H m/T}}{m^{\frac{5}{2}}} \, , \nonumber \\
a_4 & = & \frac{ d_{\alpha}}{\sqrt{2} \pi^{3/2} J^{9/2} S^{11/2}} \,
\sum^{\infty}_{n=1} \frac{e^{- \mu H n/T}}{n^{\frac{9}{2}}} \, 
- \frac{j_{\alpha}}{32 \sqrt{2} J^{9/2} S^{15/2}} \, \frac{\partial j(\mu H/T)}{\partial \mu H} \, .
\end{eqnarray}
The limit $H \to 0$ poses no problems: the spontaneous magnetization at finite temperature -- the order parameter -- is well defined. As
before, the infinite sums reduce to Riemann zeta functions, and the derivative of the three-loop function at the origin amounts to
$j'(0) = -8.8 \times 10^{-5}$ \citep{Hof11a,RPPK13}.

\begin{figure}
\begin{center}
\includegraphics[width=12.0cm]{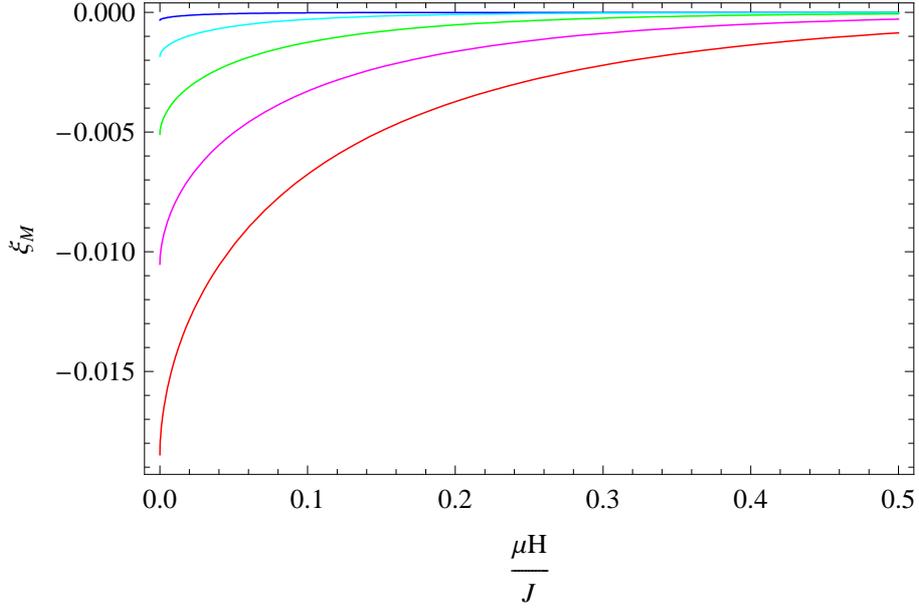}
\end{center}
\caption{[Color online] Sign and magnitude of the spin-wave interaction in the magnetization (simple cubic ideal ferromagnet,
$S=\mbox{$\frac{1}{2}$}$) as a function of the magnetic field, according to Eq.~(\ref{xiSigma}). The curves refer to the temperatures
$T/J = \{ \mbox{$\frac{1}{10}$},\mbox{$\frac{2}{10}$}, \mbox{$\frac{3}{10}$},\mbox{$\frac{4}{10}$}, \mbox{$\frac{5}{10}$} \}$ from top to
bottom in the figure.}
\label{figure6}
\end{figure}

While the sign of the interaction part in the pressure defines whether we are dealing with a repulsive or attractive interaction between
the non-ideal magnon gas particles, the sign of the interaction in the magnetization and susceptibility is related to the alignment
of the spins. In Fig.~\ref{figure6} we plot the sum of the two-loop and three-loop interaction contribution in the dimensionless quantity
\begin{equation}
\label{xiSigma}
\xi_{M}(T,H) = \frac{16 \sqrt{2} \pi^{3/2} J^{3/2} S^{5/2}}{T^{3/2}} \, \frac{M_{int}(T,H)}{S} \, ,
\end{equation}
for the five temperatures $T/J = \{ \mbox{$\frac{1}{10}$},\mbox{$\frac{2}{10}$},\mbox{$\frac{3}{10}$},\mbox{$\frac{4}{10}$},
\mbox{$\frac{5}{10}$} \}$. Again we consider the simple cubic lattice and $S=\mbox{$\frac{1}{2}$}$. The sign of $\xi_{M}$ is negative,
much like the sign of the leading Bloch term. The three-loop contribution has the same sign as the Dyson term, but this correction is very
small. The effect of the interaction is to decrease the magnetization in the entire parameter region where the effective expansion applies
($T, \mu H \ \lesssim \ 0.5 \, J$). Note that the effect of the spin-wave interaction in the spontaneous magnetization, as Dyson pointed
out a long time ago, is indeed very weak: even for $T/J = \mbox{$\frac{5}{10}$}$ and at zero magnetic field, $\xi_{M}$ is less than two
percent. 

We mention that for other systems, the behavior of the interaction part in the order parameter not necessarily follows this pattern: in
the case of the ferromagnetic (antiferromagnetic) XY model in two spatial dimensions, the interaction at finite temperature and weak
magnetic (staggered) field may increase the order parameter (see Ref.~\citep{Hof15}).

Finally, for the susceptibility,
\begin{equation}
\chi(T,H) \, = \, \frac{\partial M(T,H)}{\partial(\mu H)} \, ,
\end{equation}
we obtain the low-temperature series
\begin{equation}
\label{SusceptSeries}
\chi(T,H) = {\kappa}_0 T^{\frac{1}{2}} + {\kappa}_1 T^{\frac{3}{2}} + {\kappa}_2 T^{\frac{5}{2}} + {\kappa}_3 T^3
+ {\kappa}_4 T^{\frac{7}{2}} + {\cal O}(T^4) \, ,
\end{equation}
with coefficients ${\kappa}_i={\kappa}_i(\mu H/T)$ given by
\begin{eqnarray}
\label{SusceptCollectOrder9}
\kappa_0 & = & \frac{b_{\alpha}}{16 \sqrt{2} \pi^{3/2} J^{3/2} S^{3/2}} \, \sum^{\infty}_{n=1}
\frac{e^{- \mu H n/T}}{n^{\frac{1}{2}}} \, , \nonumber \\
\kappa_1 & = & \frac{l_{\alpha}}{\sqrt{2} \pi^{3/2} J^{5/2} S^{5/2}} \, \sum^{\infty}_{n=1}
\frac{e^{- \mu H n/T}}{n^{\frac{3}{2}}} \, , \nonumber \\
\kappa_2 & = & \frac{c_{\alpha}}{\sqrt{2} \pi^{3/2} J^{7/2} S^{7/2}} \,
\sum^{\infty}_{n=1} \frac{e^{- \mu H n/T}}{n^{\frac{5}{2}}} \, , \nonumber \\
\kappa_3 & = & \frac{2 L_{\alpha}}{\pi^3 J^4 S^5} \,
\Bigg( \sum^{\infty}_{n=1} \frac{e^{- \mu H n/T}}{n^{\frac{1}{2}}} \, \sum^{\infty}_{m=1} \frac{e^{- \mu H m/T}}{m^{\frac{5}{2}}}
+ {\Bigg\{ \sum^{\infty}_{n=1} \frac{e^{- \mu H n/T}}{n^{\frac{3}{2}}} \Bigg\}}^2 \, \Bigg) \, , \nonumber \\
\kappa_4 & = & \frac{ d_{\alpha}}{\sqrt{2} \pi^{3/2} J^{9/2} S^{9/2}} \,
\sum^{\infty}_{n=1} \frac{e^{- \mu H n/T}}{n^{\frac{7}{2}}} \,
+ \frac{j_{\alpha}}{32 \sqrt{2} J^{9/2} S^{13/2}} \,\frac{\partial^2 j(\mu H/T)}{\partial {(\mu H)}^2} \, .
\end{eqnarray}
Whereas the limit $H \to 0$ in the pressure or magnetization does not pose any problems, regarding the susceptibility, already the leading
term in the low-temperature series diverges,
\begin{equation}
\lim_{H \to 0} \kappa_0 \propto \frac{1}{\sqrt{H}} \, .
\end{equation}
It should be noted that this divergence, similar to those arising in higher-order terms, does not signal a failure of the effective field
theory framework. Rather, the singular behavior of the susceptibility is physical, as has been pointed out before \citep{SM70,KSK03}.

\begin{figure}
\begin{center}
\includegraphics[width=12.0cm]{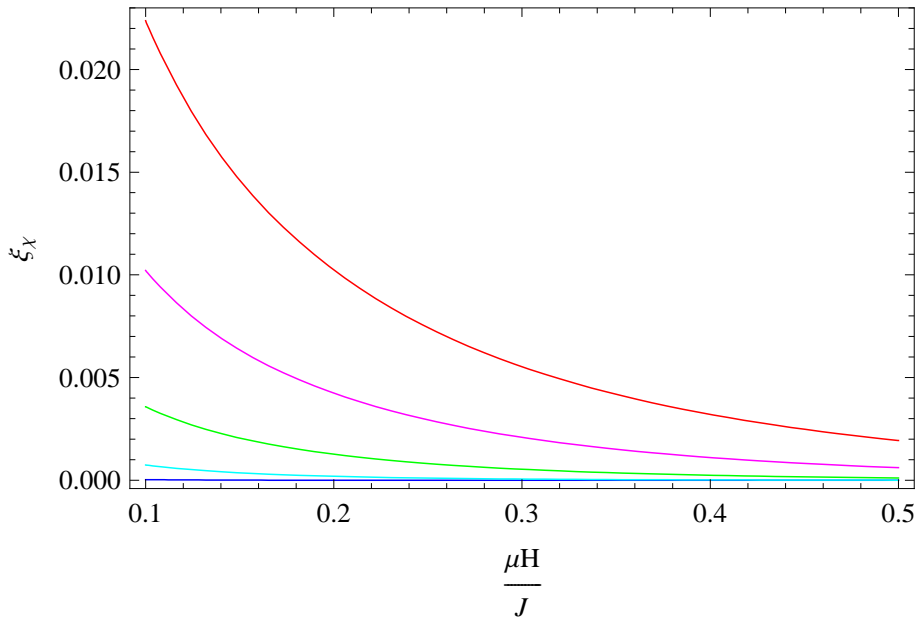}
\end{center}
\caption{[Color online] Sign and magnitude of the spin-wave interaction in the susceptibility (simple cubic ideal ferromagnet,
$S=\mbox{$\frac{1}{2}$}$) as a function of the magnetic field, according to Eq.~(\ref{xiChi}). The curves refer to the temperatures
$T/J = \{ \mbox{$\frac{1}{10}$},\mbox{$\frac{2}{10}$}, \mbox{$\frac{3}{10}$},\mbox{$\frac{4}{10}$}, \mbox{$\frac{5}{10}$} \}$ from bottom
to top in the figure.}
\label{figure7}
\end{figure}

In Fig.~\ref{figure7} we display sign and magnitude of the spin-wave interaction (sum of two-loop and three-loop contribution) in the
susceptibility, captured by
\begin{equation}
\label{xiChi}
\xi_{\chi}(T,H) = \frac{16 \sqrt{2} \pi^{3/2} J^{3/2} S^{3/2}}{T^{1/2}} \, \chi_{int}(T,H) \, ,
\end{equation}
for the simple cubic lattice, $T/J = \{ \mbox{$\frac{1}{10}$},\mbox{$\frac{2}{10}$},\mbox{$\frac{3}{10}$},\mbox{$\frac{4}{10}$},
\mbox{$\frac{5}{10}$} \}$, and $S=\mbox{$\frac{1}{2}$}$. The interaction contribution in the susceptibility is positive, featuring the
same sign as the leading Bloch term. This means that, at finite temperature, the magnetic field tends to align the spins and to enhance
the magnetization, as one would expect. Again, not all systems exhibit this behavior: in ferromagnetic (antiferromagnetic) XY models in
two spatial dimensions, the interaction contribution at finite temperatures and weak magnetic (staggered) fields in the susceptibility may
be negative \citep{Hof15}.

While we have considered the free energy density, pressure, magnetization and susceptibility, for the reader's convenience, we also
provide formulas for the energy density, entropy density, and heat capacity in appendix \ref{appendixA}. Analogous plots for these
observables confirm the above scenario: the spin-wave interaction is very weak at low temperatures, becomes most pronounced if the
magnetic field is switched off, and for a given observable does not change sign in the whole parameter regime where the effective
expansion applies. The interaction is further suppressed by the spin quantum number $S$, and manifests itself in a qualitatively analogous
manner in body-centered and face-centered cubic ideal ferromagnets.

\section{Conclusions}
\label{Conclusions}

In the present work, we have rigorously and systematically taken Dyson's program one order higher in the low-temperature expansion,
by providing explicit expressions for various thermodynamic quantities -- including free energy density, pressure, magnetization, and
susceptibility -- for the simple cubic, body-centered cubic and face-centered cubic ideal ferromagnet. In effective field theory language,
we have taken Dyson's two-loop analysis to the three-loop level. 

The low-temperature expansions involve integer and half-integer powers of $T$. The respective coefficients depend on the external magnetic
field and on -- a priori unknown -- low-energy effective constants. In our final expressions for the various observables, all low-energy
effective constants have been expressed through microscopic quantities (spin quantum number, lattice constant, exchange integral), which
is much more convenient for the condensed matter community. While the matching between effective and microscopic theory is straightforward
up to two-loop order, at the three-loop level we had to evaluate a new contribution in the microscopic theory -- fortunately related to
noninteracting spin-waves. From a conceptual point of view, we have confirmed the consistency of the effective expansion in the present
case, by showing that higher-order low-energy effective constants are successively suppressed.

One of our main themes concerned the manifestation of the spin-wave interaction in the free energy density, pressure, magnetization, and
susceptibility of cubic ideal ferromagnets at low temperatures and weak magnetic fields. Diagrammatically speaking, the interaction comes
from all partition function graphs that involve two or more loops. In the present case, the interaction shows up at order
$p^{10} \propto T^5$ (two loops) and $p^{11} \propto T^{11/2}$ (three loops) in the free energy density. In the pressure, the spin-wave
interaction turns out to be repulsive, while the sign of the interaction contribution in the magnetization (susceptibility) is negative
(positive), as one would expect.

In his pioneering articles \citep{Dys56a,Dys56b}, Dyson pointed out that the spin-wave interaction in cubic ideal ferromagnets is very
weak at low temperatures. In the present study, we have confirmed this picture: the three-loop correction that has the same sign as the
two-loop term, is very small in the entire parameter regime where the effective analysis applies ($T, \mu H \ \lesssim \ 0.5 \, J$).

The three-loop formulas provided here fully take the structure of the three cubic lattices into account. An important observation is that
the interaction contribution -- one order beyond the Dyson term -- does not involve higher-order effective constants, but only the
spontaneous magnetization at zero temperature and the spin stiffness. Both effective constants originate from the leading Lagrangian
${\cal L}^2_{eff}$ which is strictly space rotation invariant. At the three-loop level, i.e., at order $p^{11} \propto T^{11/2}$ in the free
energy density, the spin-wave interaction is thus not affected by lattice anisotropies.

Any future microscopic study carried out at the three-loop level, must end up with the result we have presented here. In this respect,
our series may be used as a reference to check efficiency and consistency of other approaches, like Green's function methods or spin-wave
theory, where spurious terms have indeed appeared in the past. On the other hand, our series may be tested against high-precision Monte
Carlo simulations of the "clean" Heisenberg model in a magnetic field, and the correctness of the effective field theory approach hence
demonstrated.

\section*{Acknowledgments}
The author would like to thank S.\ M.\ Rado\v{s}evi\'c for correspondence.

\begin{appendix}

\section{Low-temperature series for the energy density, entropy density, and heat capacity}
\label{appendixA}
\renewcommand{\theequation}{A.\arabic{equation}}
\setcounter{equation}{0}

The low-temperature series for the energy density $u$, entropy density $s$, and heat capacity $c_V$ are easily derived from the
low-temperature expansion of the pressure, using the thermodynamic relations
\begin{equation}
\label{Thermodynamics}
s = \frac{{\partial}P}{{\partial}T} \, , \quad u = Ts - P \, , \quad
c_V = \frac{{\partial}u}{{\partial}T} = T \, \frac{{\partial}s}{{\partial}T} \, .
\end{equation}
We obtain
\begin{eqnarray}
u & = &  {\cal U}_0 \, T^{\frac{5}{2}} \, + \, {\cal U}_1 \, T^{\frac{7}{2}} \, + \,  \,  {\cal U}_2 \, T^{\frac{9}{2}} \, + \, {\cal U}_3 \,
T^5 \, + \,  {\cal U}_4 \, T^{\frac{11}{2}} + {\cal O}(T^6) \, , \nonumber \\
s & = &  {\cal S}_0 \, T^{\frac{3}{2}} \, + \, {\cal S}_1 \, T^{\frac{5}{2}} \, + \,  \,  {\cal S}_2 \, T^{\frac{7}{2}} \, + \, {\cal S}_3 \,
T^4 \, + \,  {\cal S}_4 \, T^{\frac{9}{2}} + {\cal O}(T^5) \, , \nonumber \\
c_V & = & {\cal C}_0 \, T^{\frac{3}{2}} \, + \, {\cal C}_1 \, T^{\frac{5}{2}} \, + \,  \,  {\cal C}_2 \, T^{\frac{7}{2}} \, + \, {\cal C}_3 \,
T^4 \, + \,  {\cal C}_4 \, T^{\frac{9}{2}} + {\cal O}(T^5) \, . \nonumber \\
\end{eqnarray}
The quantities ${\cal U}_i={\cal U}_i(\sigma),{\cal S}_i={\cal S}_i(\sigma)$ and ${\cal C}_i={\cal C}_i(\sigma)$, with $\sigma=\mu H/T$,
in the above series are given by
\begin{eqnarray}
{\cal U}_0 & = & \frac{b_{\alpha}}{16 \sqrt{2} \pi^{3/2} J^{3/2} S^{3/2}} \,
\Bigg( \mbox{$\frac{3}{2}$} \sum^{\infty}_{n=1} \frac{e^{- \sigma n}}{n^{\frac{5}{2}}}
+ \sigma \sum^{\infty}_{n=1} \frac{e^{- \sigma n}}{n^{\frac{3}{2}}} \Bigg) \, , \nonumber \\
{\cal U}_1 & = & \frac{l_{\alpha}}{\sqrt{2} \pi^{3/2} J^{5/2} S^{5/2}} \, 
\Bigg( \mbox{$\frac{5}{2}$} \sum^{\infty}_{n=1} \frac{e^{- \sigma n}}{n^{\frac{7}{2}}} 
+ \sigma \sum^{\infty}_{n=1} \frac{e^{- \sigma n}}{n^{\frac{5}{2}}} \Bigg) \, , \nonumber \\
{\cal U}_2 & = & \frac{c_{\alpha}}{\sqrt{2} \pi^{3/2} J^{7/2} S^{7/2}} \,
\Bigg( \mbox{$\frac{7}{2}$} \sum^{\infty}_{n=1} \frac{e^{- \sigma n}}{n^{\frac{9}{2}}} 
+ \sigma \sum^{\infty}_{n=1} \frac{e^{- \sigma n}}{n^{\frac{7}{2}}} \Bigg) \, , \nonumber \\
{\cal U}_3 & = & \frac{L_{\alpha}}{\pi^3 J^4 S^5} \, 
\Bigg( 4 {\Bigg\{ \sum^{\infty}_{n=1} \frac{e^{- \sigma n}}{n^{\frac{5}{2}}} \Bigg\}}^2 
+ 2 \sigma \sum^{\infty}_{n=1} \frac{e^{- \sigma n}}{n^{\frac{3}{2}}} \sum^{\infty}_{m=1} \frac{e^{- \sigma m}}{m^{\frac{5}{2}}} \Bigg)
\, , \nonumber \\
{\cal U}_4 & = & \frac{ d_{\alpha}}{\sqrt{2} \pi^{3/2} J^{9/2} S^{9/2}} \,
\Bigg( \mbox{$\frac{9}{2}$} \sum^{\infty}_{n=1} \frac{e^{- \sigma n}}{n^{\frac{11}{2}}} 
+ \sigma \sum^{\infty}_{n=1} \frac{e^{- \sigma n}}{n^{\frac{9}{2}}} \Bigg) \, \nonumber \\
& & + \frac{j_{\alpha}}{32 \sqrt{2} J^{9/2} S^{13/2}} \, \Bigg( \mbox{$\frac{9}{2}$} \, j(\sigma) - \sigma j'(\sigma) \Bigg) \, ,
\end{eqnarray}

\begin{eqnarray}
{\cal S}_0 & = & \frac{b_{\alpha}}{16 \sqrt{2} \pi^{3/2} J^{3/2} S^{3/2}} \,
\Bigg( \mbox{$\frac{5}{2}$} \sum^{\infty}_{n=1} \frac{e^{- \sigma n}}{n^{\frac{5}{2}}}
+ \sigma \sum^{\infty}_{n=1} \frac{e^{- \sigma n}}{n^{\frac{3}{2}}} \Bigg) \, , \nonumber \\
{\cal S}_1 & = & \frac{l_{\alpha}}{\sqrt{2} \pi^{3/2} J^{5/2} S^{5/2}} \, 
\Bigg( \mbox{$\frac{7}{2}$} \sum^{\infty}_{n=1} \frac{e^{- \sigma n}}{n^{\frac{7}{2}}} 
+ \sigma \sum^{\infty}_{n=1} \frac{e^{- \sigma n}}{n^{\frac{5}{2}}} \Bigg) \, , \nonumber \\
{\cal S}_2 & = & \frac{c_{\alpha}}{\sqrt{2} \pi^{3/2} J^{7/2} S^{7/2}} \,
\Bigg( \mbox{$\frac{9}{2}$} \sum^{\infty}_{n=1} \frac{e^{- \sigma n}}{n^{\frac{9}{2}}} 
+ \sigma \sum^{\infty}_{n=1} \frac{e^{- \sigma n}}{n^{\frac{7}{2}}} \Bigg) \, , \nonumber \\
{\cal S}_3 & = & \frac{L_{\alpha}}{\pi^3 J^4 S^5} \, 
\Bigg( 5 {\Bigg\{ \sum^{\infty}_{n=1} \frac{e^{- \sigma n}}{n^{\frac{5}{2}}} \Bigg\}}^2 
+ 2 \sigma \sum^{\infty}_{n=1} \frac{e^{- \sigma n}}{n^{\frac{3}{2}}} \sum^{\infty}_{m=1} \frac{e^{- \sigma m}}{m^{\frac{5}{2}}}
\Bigg) \, , \nonumber \\
{\cal S}_4 & = & \frac{ d_{\alpha}}{\sqrt{2} \pi^{3/2} J^{9/2} S^{9/2}} \,
\Bigg( \mbox{$\frac{11}{2}$} \sum^{\infty}_{n=1} \frac{e^{- \sigma n}}{n^{\frac{11}{2}}} 
+ \sigma \sum^{\infty}_{n=1} \frac{e^{- \sigma n}}{n^{\frac{9}{2}}} \Bigg) \, \nonumber \\
& & + \frac{j_{\alpha}}{32 \sqrt{2} J^{9/2} S^{13/2}} \, \Bigg( \mbox{$\frac{11}{2}$} \, j(\sigma) - \sigma j'(\sigma) \Bigg) \, ,
\end{eqnarray}

\begin{eqnarray}
{\cal C}_0 & = & \frac{b_{\alpha}}{16 \sqrt{2} \pi^{3/2} J^{3/2} S^{3/2}} \,
\Bigg( \mbox{$\frac{15}{4}$} \sum^{\infty}_{n=1} \frac{e^{- \sigma n}}{n^{\frac{5}{2}}}
+ 3 \sigma \sum^{\infty}_{n=1} \frac{e^{- \sigma n}}{n^{\frac{3}{2}}}
+ \sigma^2 \sum^{\infty}_{n=1} \frac{e^{- \sigma n}}{n^{\frac{1}{2}}} \Bigg) \, , \nonumber \\
{\cal C}_1 & = & \frac{l_{\alpha}}{\sqrt{2} \pi^{3/2} J^{5/2} S^{5/2}} \,
\Bigg( \mbox{$\frac{35}{4}$} \sum^{\infty}_{n=1} \frac{e^{- \sigma n}}{n^{\frac{7}{2}}} 
+ 5 \sigma \sum^{\infty}_{n=1} \frac{e^{- \sigma n}}{n^{\frac{5}{2}}} 
+ \sigma^2 \sum^{\infty}_{n=1} \frac{e^{- \sigma n}}{n^{\frac{3}{2}}} \Bigg) \, , \nonumber \\
{\cal C}_2 & = & \frac{c_{\alpha}}{\sqrt{2} \pi^{3/2} J^{7/2} S^{7/2}} \,
\Bigg( \mbox{$\frac{63}{4}$} \sum^{\infty}_{n=1} \frac{e^{- \sigma n}}{n^{\frac{9}{2}}} 
+7 \sigma \sum^{\infty}_{n=1} \frac{e^{- \sigma n}}{n^{\frac{7}{2}}} 
+ \sigma^2 \sum^{\infty}_{n=1} \frac{e^{- \sigma n}}{n^{\frac{5}{2}}} \Bigg) \, , \nonumber \\
{\cal C}_3 & = & \frac{L_{\alpha}}{\pi^3 J^4 S^5} \, 
\Bigg( 20 {\Bigg\{ \sum^{\infty}_{n=1} \frac{e^{- \sigma n}}{n^{\frac{5}{2}}} \Bigg\}}^2 
+ 16 \sigma \sum^{\infty}_{n=1} \frac{e^{- \sigma n}}{n^{\frac{3}{2}}} \sum^{\infty}_{m=1} \frac{e^{- \sigma m}}{m^{\frac{5}{2}}} 
+ 2 \sigma^2 {\Bigg\{ \sum^{\infty}_{n=1} \frac{e^{- \sigma n}}{n^{\frac{3}{2}}} \Bigg\}}^2 \nonumber \\
& & + 2 \sigma^2 \sum^{\infty}_{n=1} \frac{e^{- \sigma n}}{n^{\frac{1}{2}}} \sum^{\infty}_{m=1}
\frac{e^{- \sigma m}}{m^{\frac{5}{2}}} \Bigg) \, , \nonumber \\
{\cal C}_4 & = & \frac{ d_{\alpha}}{\sqrt{2} \pi^{3/2} J^{9/2} S^{9/2}} \,
\Bigg( \mbox{$\frac{99}{4}$} \sum^{\infty}_{n=1} \frac{e^{- \sigma n}}{n^{\frac{11}{2}}} 
+ 9 \sigma \sum^{\infty}_{n=1} \frac{e^{- \sigma n}}{n^{\frac{9}{2}}} 
+ \sigma^2 \sum^{\infty}_{n=1} \frac{e^{- \sigma n}}{n^{\frac{7}{2}}} \Bigg) \, \nonumber \\
& & + \frac{j_{\alpha}}{32 \sqrt{2} J^{9/2} S^{13/2}} \, \Bigg( \mbox{$\frac{99}{4}$} \, j(\sigma) - 9 \sigma j'(\sigma)
+ \sigma^2 j''(\sigma) \Bigg) \, .
\end{eqnarray}
The coefficients $b_{\alpha}, l_{\alpha}, L_{\alpha}, c_{\alpha}, d_{\alpha}$ and $j_{\alpha}$ are listed in Eq.~(\ref{identificationSC}),
Eq.~(\ref{identificationBCC}) and Eq.~(\ref{identificationFCC}) for the tree types of cubic lattices.

\end{appendix}

\end{document}